\documentclass[aps,pra,onecolumn,notitlepage,amsmath,amssymb,showpacs,nofootinbib]{revtex4-1}
\usepackage{graphicx}
\usepackage{amsmath}
\usepackage{bm}
\usepackage{xcolor}

\newcommand{\be}{\begin{equation}}  
\newcommand{\ee}{\end{equation}}
\newcommand{\ba}{\begin{array}}
\newcommand{\ea}{\end{array}}
\newcommand{\bea}{\begin{eqnarray}}
\newcommand{\eea}{\end{eqnarray}}

\newcommand{\bra}{\langle}
\newcommand{\ket}{\rangle}

\begin{document}

\title{Roles of quantum coherences in thermal machines}

\author{C.L. Latune$^{1,2}$, I. Sinayskiy$^{1}$, F. Petruccione$^{1,2,3}$}
\affiliation{$^1$Quantum Research Group, School of Chemistry and Physics, University of KwaZulu-Natal, Durban, KwaZulu-Natal, 4001, South Africa\\
$^2$National Institute for Theoretical Physics (NITheP), KwaZulu-Natal, 4001, South Africa\\
$^3$School of Electrical Engineering, KAIST, Daejeon, 34141, Republic of Korea}

\date{\today}
\begin{abstract}
Some of the oldest and most important applications of thermodynamics are operations of refrigeration as well as production of useful energy. Part of the efforts to understand and develop thermodynamics in the quantum regime have been focusing on harnessing quantum effects to such operations. In this review we present the recent developments regarding the role of quantum coherences in the performances of thermal machines --the devices realising the above thermodynamic operations. While this is known to be an intricate subject, in part because being largely model-dependent, the review of the recent results allow us to identify some general tendencies and to suggest some future directions. 
\end{abstract}
\maketitle
\section{Introduction}
\label{intro}

Thermal machines were conceived and formalised \cite{Carnot} during the industrial revolution to produce useful energy, or {\it work}, and are still very used nowadays for that same purpose as well as for refrigeration. More than one century later, the exploration of quantum mechanics out of its initial framework, like quantum computation, quantum communication and quantum metrology, led to explore thermodynamics in the quantum regime. This was pioneered by a thermodynamic analysis of the maser \cite{Scovil_1959} and a quantum equivalence of the Carnot cycle \cite{Geusic_1967}. Since then, the emerging field of quantum thermodynamics has grown considerably, giving raise to several approaches and viewpoints. 
One commonly recurring question and very active line of research is whether quantum properties can boost the performances of quantum thermal machines \cite{Levy_2019}.
In this perspective, we review in the following recent developments on the role of quantum coherences -one essential characteristic of quantum states- in the performances of quantum thermal machines.

\begin{figure}\centering
\resizebox{0.6\columnwidth}{!}{  \includegraphics{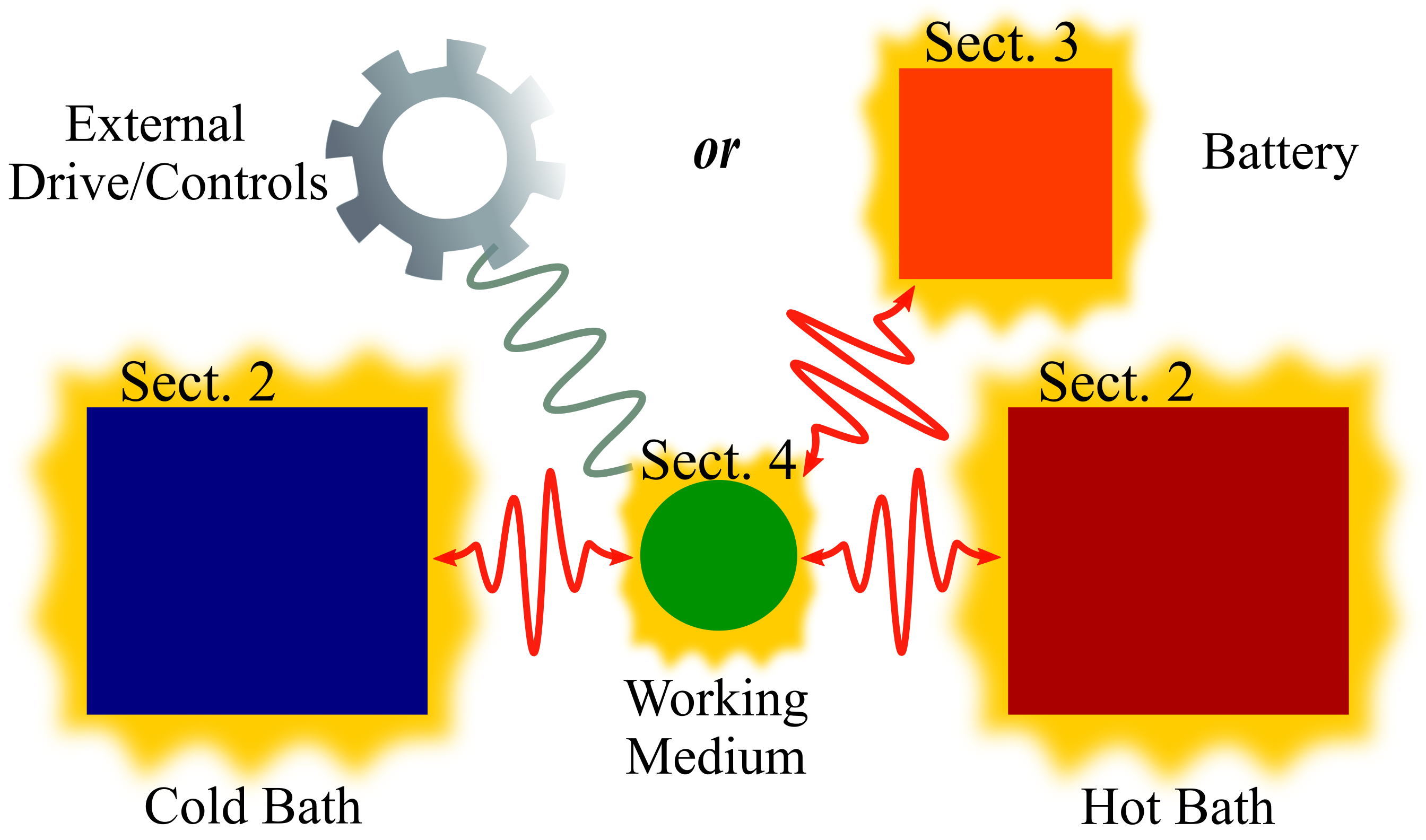} }
\caption{Illustration of the different situations
 considered in this review: coherences contained in the baths (Sect.~\ref{secbath}), coherences contained in the quantum battery (Sect.~\ref{secbattery}), coherences in the working medium (Sect.~\ref{secwm}), and briefly non-local coherences (Sect.~\ref{secnl})(not illustrated). Coherences are represented by a yellow crown surrounding a system. 
 }
\label{review}       
\end{figure}

 Before starting, we recall very briefly the basic features of the main designs of thermal machines to fix the nomenclature. Thermal machines are composed of a cold and a hot thermal baths which interact with a core system called working medium as in Fig.~\ref{review}. When this working medium is a quantum system, the thermal machine is usually qualified as ``quantum'' as well. A machine can operate cyclically thanks to external controls switching on and off the contact with the thermal baths.    
 %
%
 We consider the two mostly used cyclic designs. The first one, the quantum Carnot cycle \cite{Geusic_1967,Quan_2007}, adapted from the famous Carnot cycle \cite{Carnot}, is a four-stroke cycle containing an isentropic compression, an isothermal expansion in contact with the hot bath, then an isentropic expansion, and an isothermal compression in contact with the cold bath, closing the cycle. 
The second cycle is the quantum Otto cycle \cite{Quan_2007}, consisting also in four strokes: an isentropic compression, an isochore in contact with the hot bath, an isentropic expansion, and finally an isochore in contact with the cold bath. When the cycles are realised in the above order, the thermal machine is an engine, extracting work from the baths. When the cycles are reversed, the machines becomes refrigerators. 
Contrasting with these cyclic machines, continuous thermal machines \cite{Kosloff_2014} are continuously in contact with both hot and cold baths, and can be externally driven, or operate autonomously thanks to a ``battery'' (see Fig.~\ref{review}) which can be a third thermal bath \cite{Mitchison_2019}, sometimes called work bath, or a single quantum system \cite{Gelbwaser_2014}. Continuous machines operate as engines or refrigerators depending on the initial conditions.



 Throughout the review, quantum coherences mean non-diagonal elements of a density operator describing the state of a quantum system (illustrated by the yellow crown in Fig.~\ref{review}). Quantum coherences are basis-dependent. 
 In this review, all coherences are implicitly defined in the natural eigenbasis of local Hamiltonians. 
 The question of whether quantum coherences can enhance thermal machines performances is still highly debated. This is in part because it is largely model dependent. In order to try to classify the diverse effects and to identify some general tendencies, 
we will divide the appearance of quantum coherences --simply called coherences in the following-- in several categories, see Fig.~\ref{review}. Firstly, we will analyse coherences present in the baths (Sect.~\ref{secbath}), and then coherences present in the battery (Sect.~\ref{secbattery}). Secondly, Sect.~\ref{secwm} is dedicated to review some consequences of the presence of coherences in the working medium. This large section is divided accordingly to the mechanism responsible for the introduction of coherences in the working medium. Finally, Sect.~\ref{secnl} briefly mentions non-local coherences between the system and the baths. 
Some final remarks and conclusions are drawn in the recapping Sect.~\ref{concl}.

\section{Quantum coherences in the baths}\label{secbath}


 While thermal machines were initially designed to operate with thermal baths, it is quite natural to extend to baths containing quantum properties and in particular coherences, Fig.~\ref{review}. Several studies considered this question. We separate them in two categories: bath containing coherences between levels of different energy and baths containing coherences between levels of same energy. 

 \subsection{Baths containing coherences between levels of different energy}\label{secbathvcoh} 
 One of the most studied example of baths containing coherences between levels of different energy, called {\it vertical coherences} in the following, is squeezed baths \cite{Huang_2012,Abah_2014,RoBnagel_2014,Manzano_2016,Niedenzu_2016,Niedenzu_2018,Xiao_2018,Alicki_2015,Agarwalla_2017}. Such baths consist of an ensemble of modes or harmonic oscillators in squeezed states. More precisely, denoting by $H_B = \sum_k \omega_k b_b^{\dag}b_k$ the free Hamiltonian of the bath, where $\omega_k$, $b_k$, and $b_k^{\dag}$ are respectively the energy, annihilation operator, and creation operator of the mode $k$, a squeezed state is described by the density operator of the form $\rho_B(\xi,T)= S(\xi)\rho^{\rm th}(T)S^{\dag}(\xi)$ where $\rho^{\rm th}(T)$ is a thermal state at temperature $T$, $S(\xi)= \otimes_k e^{\frac{\xi}{2} (b_k^{\dag})^2-\frac{\xi^{*}}{2}b_k^{2}}$ 
 is the squeezing operator, and $r=|\xi|$ and $\phi={\rm arg} \xi$ are respectively the squeezing factor and the phase of the squeezing \cite{Ferraro_2005}. In particular, a squeezed state contains vertical coherences between any levels separated by a pair number of excitations: $_k\langle n + 2q|\rho_B(\xi,T)|n\rangle_k \ne 0$, for all positive integer $n$ and $q$, where $|n\rangle_k$ are the Fock state of the mode $k$.

Coming back to thermal machines, considering a cold bath in a thermal state at temperature $T_c$ and a hot bath in a squeezed state $\rho(\xi, T_h)$, it was shown for Otto engines \cite{Huang_2012,Abah_2014,RoBnagel_2014,Manzano_2016,Niedenzu_2016,Niedenzu_2018,Xiao_2018} and continuous engines \cite{Alicki_2015,Agarwalla_2017} that the maximal achievable efficiency is, in the limit of high $T_h$, 
\be
\eta_{\rm max} = 1- \frac{T_c}{T_h(1+2{\rm sinh}^2r)}, 
\ee
which is strictly larger than the Carnot efficiency as long as $r>0$. Additionally, the efficiency at maximal power becomes $\eta^{*}=1-\sqrt{\frac{T_c}{T_h(1+2{\rm sinh}^2r)}}$, also strictly larger than the Curzon-Ahlborn bound \cite{CAbound} as soon as $r>0$. These results were verified experimentally with a Otto engine in \cite{Klaers_2017}.

Although the performances are actually enhanced and the hot bath contains coherences there is no relation between both. The efficiency increases stem simply from the increased energy brought by the squeezing of the hot bath \cite{paperautonomous}, which can also be seen as a consequence of the non-thermal distribution of the populations. In particular, dephased squeezed states \cite{Alicki_2014} yield the same efficiencies. This can be seen using the concept of apparent temperature \cite{paperapptemp,heatflowreversal}, also called local temperature in \cite{Alicki_2014,Alicki_2015}, briefly introduced in the following. For a system $S$ interacting with a bath $B$, both in arbitrary states, and assuming the usual Born, Markov, and secular approximations \cite{Petruccione_Book}, particularly well-suited for the study of thermodynamics in the quantum regime, the direction of the heat flow between $S$ and $B$ is determined by their {\it apparent temperatures}, more precisely is from the hottest apparent temperature to the coldest. The apparent temperatures are defined by \cite{paperapptemp}
\be
{\cal T}_x(\omega)= \omega \left( \ln \frac{\bra A_x(\omega)A_x^{\dag}(\omega)\ket_{\rho_x}}{\bra A_x^{\dag}(\omega)A_x(\omega)\ket_{\rho_x}}\right)^{-1},
\ee
where $\bra O\ket_{\rho_x}:={\rm Tr}_x \rho_x O$ denotes the expectation value of the operator $O$ taken in the state represented by the density operator $\rho_x$, $x=S,B$, and $A_x(\omega)$ are the eigenoperators associated to the operators $A_x$ involved in the interaction between $S$ and $B$. In particular, $A_S(\omega)$ corresponds to the jump operators appearing in the Markovian master equation describing the reduced dynamics of $S$, and $\omega$ is the frequency transition associated to each ``jump''. 

Therefore, for an engine operating with a hot bath in the squeezed state $\rho_h(\xi,T_h)$, the engine does not ``see'' the temperature $T_h$ but the apparent temperature ${\cal T}_h =\omega \left(\ln\frac{{\rm tanh}^2r+e^{\omega/T_h}}{e^{\omega/T_h}{\rm tanh}^2r+1}\right)^{-1}$ \cite{paperautonomous,Alicki_2015}, where $r$ is the squeezing factor introduced above and $\omega$ is the frequency transition of the working medium (most designs, if not all, have only one frequency transition involved in the coupling with each bath).
Consequently, the maximal achievable efficiency is not given by the usual Carnot bound using $T_h$, but by $\eta \leq \eta_{\rm ext C}:= 1-\frac{T_c}{{\cal T}_h} $ which is an extended Carnot bound obtained using the apparent temperature ${\cal T}_h$. In the limit of high temperature $k_BT_h\gg \hbar\omega$, ${\cal T}_h \simeq T_h(2{\rm sinh}^2r+1)$ and one recovers the expression of $\eta_{\rm max}$ mentioned above. In the same way, this is valid for the Curzon-Ahlborn bound which can be extended using apparent temperatures.

 Since vertical coherences do not affect the apparent temperatures \cite{paperapptemp,heatflowreversal}, the vertical coherences present in the squeezed hot bath are not responsible for the above efficiency increases. Additionally, one can see that the hot bath apparent temperature is entirely determined by its energy so that the real reason for the efficiency increases is the increased energy of the squeezed bath.

Finally, for autonomous refrigerators operating with a cold bath at $T_c$, a hot bath at $T_h$ and a ``work bath'' (playing the role of the battery) in a squeezed state $\rho_{w}(\xi,T_w)$, the same phenomenon takes place: the working medium sees the work bath at its apparent temperature ${\cal T}_w=\omega \left(\ln\frac{{\rm tanh}^2r+e^{\omega/T_w}}{e^{\omega/T_w}{\rm tanh}^2r+1}\right)^{-1}$. Consequently, the maximal achievable efficiency must be $\eta_{\rm refri, max} = \frac{T_c}{T_h-T_c}\left(1-\frac{T_h}{{\cal T}_w}\right)$ (where ${\cal T}_w$ substitutes $T_w$) as directly shown in \cite{Correa_2014}. Hence, the above conclusion on the role of the vertical coherences is also valid in this situation. \\

 Beyond squeezed baths, some studies investigate general coherent baths \cite{Dag_2016,Rodrigues_2019}. The consequence of horizontal coherences is the apparition of a unitary contribution in the reduced dynamics of the working medium. Note that because of the peculiar structure of squeezed states (containing coherences only between second neighbours, $_k\langle n + 2p|\rho_B(\xi,T)|n\rangle_k$), such unitary contribution cancels out. 
The energy exchanges associated to such unitary contribution can be identified as work. 
However, one major issue with vertical coherences in baths is the lack of tangible perspectives for realisations of coherent baths.
This is due to the problem of the phase of the coherences \cite{paperapptemp}. Phases can be thought as clocks. To interfere constructively, they need to be prepared in phase otherwise the coherences will average to zero. Since a bath typically contains a macroscopic number of particles or modes, it becomes practically impossible to prepare all these subsystems with the same phases. 
The collisional model greatly relaxes these experimental constraints but still represents a huge technical challenge.  
Additionally, we recall that a state containing vertical coherences $\rho_B$ is not a stationary state ($[H_B,\rho_B]\ne0$). Therefore, the instant of time at which the working meduim starts interacting with the bath is crucial. How is this time defined? Can it be controlled? 
Such issues disappear for horizontal coherences.

\subsection{Baths containing coherences between levels of same energy}\label{secbathhcoh}
 In the following we will refer to coherences between levels of same energy as {\it horizontal coherences}. 
 Baths containing horizontal coherences induce a reduced dynamics of the working medium of the same form as thermal bath \cite{Alicki_2014,Alicki_2015,paperautonomous}. The only difference is that the damping rates are characterised by the apparent temperatures \cite{paperapptemp} of the baths. In other words, as in the previous section, the working medium only ``sees'' the apparent temperatures of the baths.
 Some example of baths containing horizontal coherences includes the phaseonium \cite{Scully_2003}, which is a three-level atoms containing coherences between the two near degenerate ground states, multi phaseonium (two-level system with multiple degenerate ground state) \cite{Turkpence_2016}, extensions to finite-time phaseonium-based engine \cite{Guff_2019}, and thermally entangled pair of two-level atoms \cite{Dillenschneider_2009}. More general frameworks for arbitrary stationary baths (such that $[\rho_B,H_B]=0$) were also established  \cite{Niedenzu_2016,Alicki_2015,Liberato_2011,Li_2014}. Note that although the systems used in \cite{Scully_2003,Turkpence_2016,Guff_2019,Dillenschneider_2009} do not constitute baths per se, they can reproduce bath properties through collisional model \cite{paperapptemp,Filipowicz_1986,Milburn_1987,Alicki_1987}, which consists in fast sequences of interaction with the working medium followed by a re-initialisation.

 The efficiency enhancements reported in the above studies were associated to diverse phenomena, from coherences to degeneracy and entanglement. However, it was pointed out in \cite{Li_2014} that the common cause to such enhancements was coherences. This can be refined by saying that the enhancements come from the contribution of the horizontal coherences to the baths' apparent temperatures. Moreover, the actual efficiencies can be obtained straightforwardly using the bath apparent temperatures in the place of bath temperatures. 
More precisely, decomposing the bath's state into $\rho_B=\rho_{\rm pop} + \chi$, with $\rho_{\rm pop}$ containing only the populations and $\chi$ only the horizontal coherences, its apparent temperature, as seen from the working medium's viewpoint, can be expressed as 
\cite{paperapptemp,heatflowreversal}:
\be
\frac{\omega}{{\cal T}_{\rm bath}} = \frac{\omega}{{\cal T}_{\rm pop}} + \ln \frac{1+c_+}{1+c_{-}}
\ee
where $\omega$ is the bath frequency transition in resonance with the working medium, ${\cal T}_{\rm pop} =\omega \left(\ln\frac{\bra A_{B}(\omega)A_B^{\dag}(\omega)\ket_{\rho_{\rm pop}}}{\bra A_{B}^{\dag}(\omega)A_B(\omega)\ket_{\rho_{\rm pop}}}\right)^{-1}$ is a contribution exclusively from the populations, while $c_+ = \frac{\bra A_B(\omega)A^{\dag}_B(\omega)\ket_{\chi}}{\bra A_B(\omega) A^{\dag}_B(\omega)\ket_{\rho_{\rm pop}}}$ and $c_{-}=\frac{\bra A^{\dag}_B(\omega) A_B(\omega) \ket_{\chi}}{\bra A^{\dag}_B(\omega) A_B(\omega)\ket_{\rho_{\rm pop}}}$ are contributions from horizontal coherences. Note that if the populations are thermally distributed according to a temperature $T_B$, then ${\cal T}_{\rm pop} =T_B$.
 Additionally, coherences make the apparent temperature ``hotter'', meaning $\big|\frac{1}{{\cal T}_B}\big|<\big|\frac{1}{{\cal T}_{\rm pop}}\big|$, if and only if $\bra A_B(\omega)A^{\dag}_B(\omega)\ket_{\chi}<\bra A_B^{\dag}(\omega)A_B(\omega)\ket_{\chi}$. Then, if $B$ plays the role of the hot (cold) bath, horizontal coherences are beneficial if and only if $\bra A_B(\omega)A^{\dag}_B(\omega)\ket_{\chi}<\bra A_B^{\dag}(\omega)A_B(\omega)\ket_{\chi}$ ($\bra A_B(\omega)A^{\dag}_B(\omega)\ket_{\chi}>\bra A_B^{\dag}(\omega)A_B(\omega)\ket_{\chi}$).
 For instance, in the situation of the phaseonium \cite{Scully_2003}, one obtains that the coherences between the degenerate ground states increase the efficiency of the engine if and only if their real part is negative, recovering the conclusion of \cite{Scully_2003}, and as already mentioned the actual efficiency is obtained using the bath apparent temperature instead of place of the temperature.

\section{Coherences in the quantum battery}\label{secbattery} 
Pioneering studies of autonomous thermal machines powered by single or small quantum systems (not baths) analysed the influence of the battery's initial state \cite{Gelbwaser_2014,Gelbwaser_2015} but without focusing specifically on coherences.  
 In a similar but more general setup, relying on dispersive coupling between battery and working medium (shown to be likely the only viable two-body coupling), the impact of coherences in batteries powering autonomous machines is investigated \cite{paperautonomous}, Fig.~\ref{review}. It is shown that the maximal achievable efficiency is simply characterised by the apparent temperature \cite{paperapptemp,heatflowreversal} of the battery. As already mentioned in the previous section, the apparent temperature is sensitive only to horizontal coherences, and the same criteria about positive or negative impact on the apparent temperature and efficiency hold \cite{paperapptemp}.

  Additionally, the interaction of the engine with the battery may generate correlations between engine and battery \cite{Doyeux_2016} but also coherences within the battery. For instance, in a Otto engine put in contact with a quantum battery (a single harmonic oscillator), it is shown \cite{Watanabe_2017} that due to the accumulation over cycles of coherences within the battery, the amount of work loaded in the battery is not proportional to the number of applied cycles, contrasting with the work loaded when coherences are discarded (by projective energy measurements after each cycle). Moreover, this manifestation of quantum coherences is shown to have no classical counterpart and to yield significant increase of extracted work. 
  
  Conversely, the interaction engine-battery can generate coherences within the working medium. 
  In particular, one promising setup is an ensemble of engines collectively coupled to the same battery. The work stored in the battery can be enhanced thanks to collective effects \cite{Watanabe_2020}, but the precise role of the battery-induced coherences remains to be further investigated.
 


\section{Coherences within the working medium}\label{secwm}
Differently from coherences in the baths or in the battery, coherences within the working medium do not need to be added externally since it can be generated naturally by the dynamics of the machine itself. This also provides a fairer comparison between classical and quantum machines. In the following we classify the coherences within the working medium accordingly to the mechanism behind their generation.

\subsection{Drive-induced coherences}\label{secnonadiab}
Both cyclic and continuous machines relies on external driving of the working medium, described by a time-dependent Hamiltonian $H_{\rm WM}(t)$. Coherences within the working medium is generated by the external drive when it stops being adiabatic (in the quantum mechanical sense \cite{Teufel_2003,Allahverdyan_2005,Albash_2012}), which means typically rapid modulations of the drive and $[H_{WM}(t),H_{WM}(t')]\ne0$. 
One can show in simple unitary or isothermal processes that non-adiabatic drive and the subsequent generation of coherences give raise to extra energetic cost when compared to adiabatic driving \cite{Plastina_2014}. These extra energetic costs are sometimes referred to as {\it irreversible work} (for isolated systems) or {\it inner friction} (for isothermal processes), and can be understood remembering that work can be extracted from coherences \cite{Korzekwa_2016}. Therefore, the generation of coherences must have some energetic costs. Note that even in situations where coherences are naturally generated by the bath (as in Section \ref{secbathinduced}) it was shown that there is still an energetic cost \cite{negentropyprod}. 
Consequently, drive-induced coherences generated during an isentropic stroke and damped into one of the thermal bath during the subsequent isochore are wasted resource which inevitably leads to a reduction of the performances of the machines. This has been shown explicitly first in cyclic engines operating with ensembles of interacting spins $1/2$ \cite{Kosloff_2002,Feldmann_2003,Feldmann_2004}, leading to the introduction of the concept of {\it quantum friction}, and later generalised to arbitrary working systems \cite{Plastina_2014,Brandner_2016,Brandner_2017,Brandner_2020}, . 

The simplest way to avoid such extra energetic costs is to realise drives which are slow enough \cite{Teufel_2003,Allahverdyan_2005,Albash_2012}. However, very long strokes will considerably reduce the output power of engines and refrigerators. Alternative strategies have been elaborated in order to avoid long strokes while still limiting quantum friction. The first one, called quantum lubrification \cite{Feldmann_2006}, aims at suppressing the generation of coherences during the external driving. This can be realised by adding random noise in the implementation of the driving process in order to synthesise a dephasing dynamics \cite{Feldmann_2006}. Further strategies also relying on additional external controls have been established and regrouped under the umbrella term {\it shortcut to adiabaticity} \cite{Berry_2009,Chen_2010} with experimental implementations \cite{Deng_2018} and applications in Otto \cite{delCampo_2014,Deng_2013,Beau_2016,Abah_2019,Hartmann_2020} and Carnot \cite{Dann_2020} engines.
 Such techniques inevitably require injection of extra energy \cite{Campbell_2017,Abah_2019b,Abah_2018} since relying on additional external controls, but the overall energy balance indicates that shortcut to adiabaticity can still be beneficial \cite{Abah_2018,Abah_2019c}.  
 

Instead of trying to reduce the non-adiabadicity and its inherent energy loss, an interesting alternative \cite{Uzdin_2016} to short-cut to adiabaticity consists in using its by-products, namely coherences. For instance, such coherences can be extracted and adequately injected in other thermal machines, boosting their output power \cite{Uzdin_2016}.  

In the same spirit, in order to avoid damping the coherences in the thermal baths just after their generation by non-adiabatic drives, one can maintain them (by reducing the isochoric strokes) and see if they could be useful in the next isentropic stroke. This is the object of the next section. 


\subsection{Coherences maintained throughout the whole cycle}\label{secconservedcoh}


In Otto engines or refrigerators, coherences generated during an isentropic stroke by non-adiabatic drive $[H_{\rm WM}(t),H_{\rm WM}(t')]\ne 0$ can be partially maintained until the next isentropic stroke if the preceding isochore is stopped before the working medium reaches full thermalisation \cite{Dann_2020,Feldmann_2012,Camati_2019,Altintas_2015,Uzdin_2015}. Such kind of ``incomplete'' isochores enable interesting possibilities of interplay and interferences during the subsequent isentropic expansion \cite{Camati_2019}. 
Indeed, when coherences are maintained throughout the cycles, efficiency and power enhancements were shown to be possible in Otto engines operating with a two-level system strongly coupled to a harmonic oscillator as working medium \cite{Altintas_2015} as well as in Carnot cylces \cite{Dann_2020}, while one would have expected reduction of performances due to quantum friction.

More precisely, it can be shown without particular model for the working medium that the amount of coherences preserved during the isochore can reduce the quantum friction emerging during the isentropic strokes \cite{Camati_2019}. 
 However, as often with coherences in thermal machines, beneficial effects are not systematic. Depending on the duration of each stroke, the maintained coherences can either have positive or negative effects on the output power and efficiency of the engine when compared with the dephased engine (coherences discarded at the end of the incomplete isochores) \cite{Camati_2019}. 
 
Additionally, power increase can be associated to a mechanism of coherent work extraction \cite{Uzdin_2015}. Such coherent work extraction is also present in continuous engines, and was even shown to be necessary \cite{Uzdin_2015}, and experimentally confirmed \cite{Klatzow_2019}. However, this necessity of coherent work extraction, and therefore of coherences, seems to be model-dependent so that further studies would be welcomed in continuous engines.

 \subsection{Bath-induced coherences}\label{secbathinduced}
%
The perspective of enhancements stemming from bath-induced coherences is particularly appealing since it comes somehow for free \cite{Scully_2011,Svidzinsky_2012}: no extra energy is invested, no extra resource is used, and no need to inject coherences as in Section \ref{secbath}. However, as we will see, it is not so simple. There is an ``experimental price'' to pay.

\subsubsection{Bath-induced coherences in cyclic many-body thermal machines}\label{seccyclicmb}
Several positive effects can be obtained from bath-induced coherences in many-body working medium. 
 Firstly, ensemble of qubits or spins interacting collectively with a thermal bath can reach equilibrium much faster than through individual coupling of each qubit with the bath. This fast equilibration can be used to reduce the duration of the isochoric strokes and therefore increase the power of cyclic engines \cite{Kloc_2019,colheatcapacity}. Note that such fast equilibration is a direct consequence of the collective coupling with the bath which inherently generates coherences in the ensemble (see more details below). However, if one cancels such coherences, the speed-up in the equilibration is also cancelled. Additionally, although both rely on collective bath coupling, the fast equilibration is different from superradiance \cite{Gross_1982} since not directly related to energy rate and not limited to half excited states.

 Beyond this ``super-equilibration'' boost, some steady state phenomena also associated to bath-induced coherences can bring beneficial effects.
%
%
%
%
Consider an ensemble of $n$ spins of arbitrary dimension $s$, coupled to a thermal bath through a collective coupling of the form $V= J_x O_B$, where $J_x= \sum_{k=1}^n j_{x,k}$ is the x-component of the collective spin operator, sum of the local spin operators $j_{x,k}$, and $O_B$ is a bath operator. 
This collective dynamics is compared with the independent one, characterised by a coupling of the form $V= \sum_{k=1}^n j_{x,k}O_{B,k}$, where each bath operator is independent from each other, in the sense that ${\rm Tr}_B \rho_B O_{B,k}O_{B,k'} = 0$ if $k\ne k'$. Physically, it happens when the spins are distinguishable from the point of view of the bath \cite{Gross_1982,bathinducedcoh,bathinducedcohTLS}, which includes situations where the spins have perpendicular dipole transitions and/or are far apart (compared to the characteristic length-scale of the bath, see more details in \cite{Gross_1982}). This indistinguishability requirement constitutes an experimental challenge which might be seen as the ``price to pay'' for the enhancements discussed in the follwoing.

\begin{figure}\centering
\resizebox{0.65\columnwidth}{!}{  \includegraphics{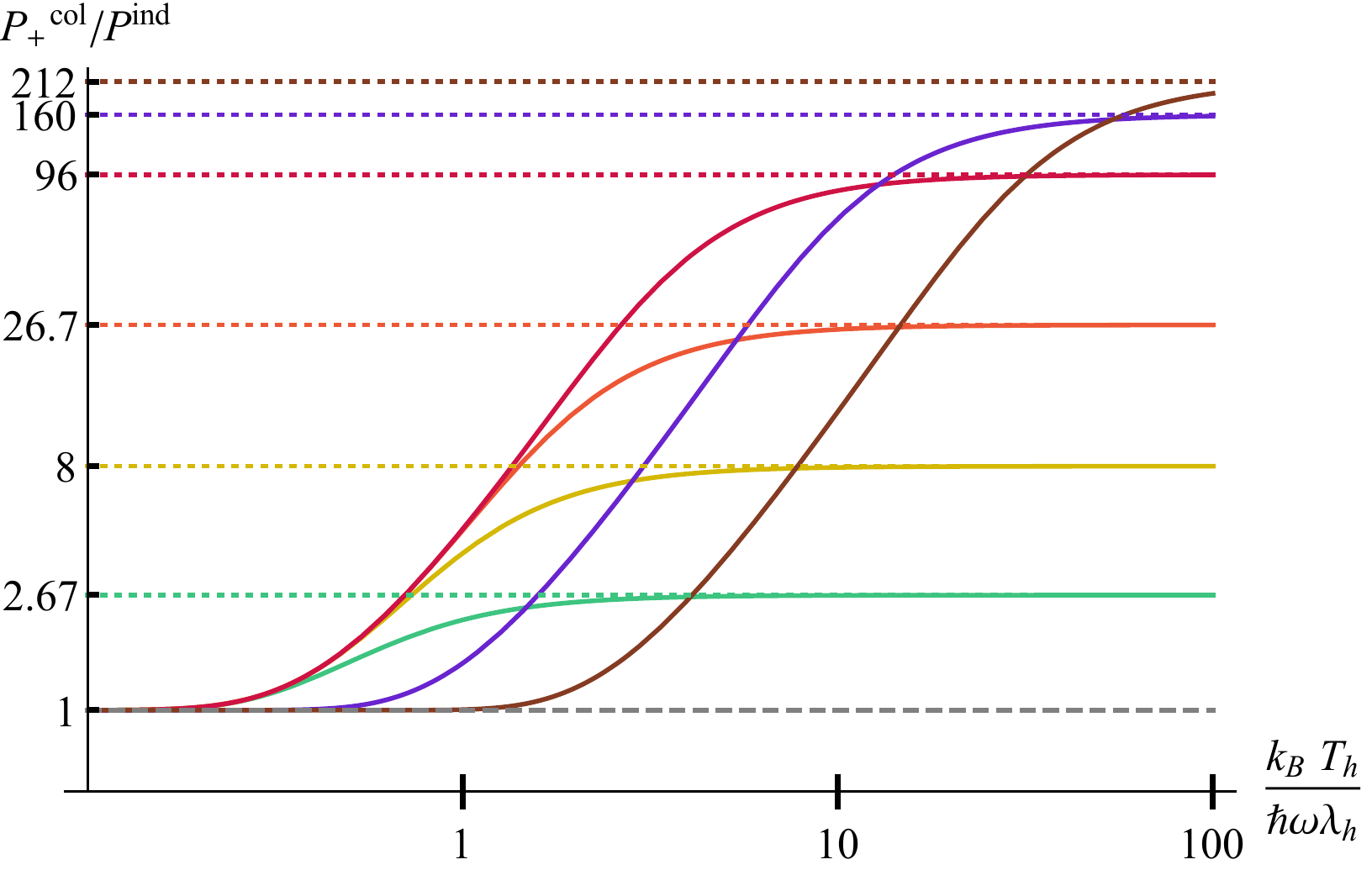} }
\caption{Plot in log-log scale of the ratio of the collective engine output power, $P_{+}^{\rm col}$, by the independent engine output power, $P^{\rm ind}$, in function of $\frac{k_BT_h}{\hbar\omega\lambda_h}$. The curves correspond to ensembles containing $n=2$ (green curve), $n=4$ (yellow curve), $n=8$ (orange curve), $n=16$ (pink curve) of spins $s=1/2$. The purple and brown curve correspond to an ensemble of $n=16$ spins $s=3/2$ and $s=9/2$, respectively. The horizontal dashed lines indicate the asymptotic value given by $\frac{n(ns+1)}{s+1}$. }
\label{ratiopowers}       
\end{figure}

One consequence of the collective coupling is that the induced dissipation, described by a master equation derived under Born, Markov, and secular approximations, occurs through collective jump operators $J_{\pm}= \sum_{k=1}^n j_{\pm,k}$ sum of the local jump operators $j_{\pm,k} = j_{x,k} \pm i j_{y,k}$. Such collective jump operators represents the collective absorption and emission of excitations from or to the bath, generating excitations delocalised between the spins. These collective excitations are the mechanism behind the well-known bath-induced coherences present for instance in superradiance \cite{Gross_1982}, bath-induced entanglement \cite{Kim_2002,Plenio_2002,Benatti_2003} and recently shown to reduce the entropy production \cite{negentropyprod}. 
Additionally, the bath-induced coherences happen to be ``horizontal'' (coherences between levels of same energy), hence affecting the apparent temperature of the spin ensemble, see \cite{paperapptemp} and also Sect.~\ref{secbath}. As a consequence, the steady state, and more importantly its energy, are significantly affected, giving raise to a phenomenon of mitigation of the bath's action \cite{bathinducedcohTLS,bathinducedcoh}. More precisely, for a spin ensemble initially in a thermal state at temperature $T_0$ and interacting collectively with a thermal bath at temperature $T_B > T_0$ ($T_B<T_0$), the steady state energy $E^{\rm ss}$ reached by the spin ensemble is strictly lower (larger) than the thermal equilibrium energy $E^{\rm th}(T_B)$ (reached when each spin of the ensemble interacts independently with the bath) .  
Strengthening the idea that collective coupling mitigates the action of the bath, it was also shown in \cite{bathinducedcoh} that for any initial temperature of the spin ensemble, its free energy variation $|\Delta {\cal F}|$ during the process of collective dissipation is reduced (in absolute value) compared to the free energy variation during independent dissipation. 

Coming back to thermal machines, pursuing their performance enhancements suggests that one should look for amplifying the action of the baths rather than mitigating it. However, combining the mitigation effects from two baths at different temperatures, one can achieved an overall amplification of the baths' action. Applied to Otto engines, significant power increase can be obtained when compared to engines made of spins interacting independently with the thermal baths \cite{bathinducedcoh}, referred to as ``independent engines'' in the following. Importantly, the power increase is not systematic and happens for certain ranges of cold and hot bath temperatures \cite{bathinducedcoh} but becomes larger and larger as the hot bath temperature goes to infinity. The most favourable situation is when the spin ensemble's initial state belongs to the symmetrical subspace \cite{Gross_1982} (also called Dick subspace), which includes thermal states with extremal temperatures, meaning $\hbar\omega/k_B|T_0|\gg1$, where $\hbar\omega$ is the energy transition of the spins and $k_B$ the Boltzmann constant. In such situation, the extracted work by the collective engine can be up to $(ns+1)/(s+1)$ times larger than the work extracted by the independent engine, for asymptotic hot and cold bath temperatures $\hbar\omega/k_BT_c\gg1$ and $\hbar\omega/k_BT_h\ll1$. These conditions are not necessarily achievable in practice but it gives an idea of the possible extent of the enhancement brought by collective coupling. 
 Note also that the efficiency remains unaffected by these collective effects.
 
These results were confirmed and extended for Otto engines operating close to the Carnot efficiency \cite{colheatcapacity} by introducing collective heat capacity. In such regime, $\lambda_h/T_h \simeq \lambda_c/T_c$, where $\lambda_c$ and $\lambda_h$ are respectively the expansion and compression factors of the Otto cycle, with $\lambda_h>\lambda_c$. 
Again, in the most favourable situation when $\hbar\omega/k_B|T_0|\gg1$ and for high temperature of the hot bath ($k_B T_h/\hbar\omega\lambda_h\gg 1$), the work extracted per cycle by the collective engine is multiplied by $(ns+1)/(s+1)$ with respect to the independent engine, coinciding with what just seen in the above paragraph. 
Taking into account the collective fast equilibration \cite{Kloc_2019,colheatcapacity} mentioned in the beginning of this section, the duration of the collective isochoric strokes can be divided by $n$, resulting in an output power multiplied by $n$. 
With this further improvement, the collective engine is always (for any bath temperature) performing as well as or better than the independent engine, see Fig. \ref{ratiopowers}. For high hot bath temperatures, the collective coupling yields an $n(ns+1)/(s+1)$-fold enhancements with respect to independent engines (see Fig. \ref{ratiopowers}).


 In conclusion of this long section, the above enhancements rely on horizontal coherences induced by the bath thanks to the collective coupling. Indeed, if one suppresses these coherences, the collective dissipation leads to a unique steady state which is the thermal equilibrium state \cite{bathinducedcohTLS,bathinducedcoh}, as for independent dissipation. Therefore, without bath-induced coherences all the above phenomena and enhancements would not happen. Nevertheless, alternative viewpoints \cite{Niedenzu_2018b,Holubec_2018,Holubec_2019} emphasise that the problem can be described using the Dick states \cite{Gross_1982}  (only when $\hbar\omega/k_B|T_0|\gg1$) for which the spin ensemble's dynamics is equivalent to the one of a single spin of dimension $ns$, removing all the coherences from the problem. Still, this suggests a comparison between two different systems. Additionally, single systems with spin $ns$ are not always accessible (especially for large $n$). 
  We come back on this point in the conclusion.   


\subsubsection{Bath-induced coherences in continuous many-body thermal machines}\label{secmb}
The continuous counterpart of the cyclic engines \cite{bathinducedcoh,colheatcapacity} described in the previous section is analysed in \cite{Niedenzu_2018b} using spins $s=1/2$. 
 The external driving consists in a periodic energy modulations of each spins, $H_{\rm WM} =\hbar \frac{\omega(t)}{2} \sum_{k=1}^n \sigma_{z}^{k}$. 
As in the previous section, the collective engines, where all spins interact collectively with the baths, and the independent engines, characterised by independent spin-bath interactions, are compared. 
It is shown that the output power of the collective engine can be up to $(n+2)/3$ times larger than the independent one \cite{Niedenzu_2018b}. Such enhancement stems from a kind of ``steady-state superradiance''. It is reached when the spin ensemble's state belongs initially to the symmetrical subspace \cite{Gross_1982} and when the effective temperature characterising the steady state of the spin ensemble tends to infinity, so that the steady state is indeed a superradiant state \cite{Gross_1982}. This effective temperature depends on the bath temperatures, bath spectral densities, and driving frequency. 
Conversely, at low effective temperature the collective spin engine exhibits the same output power as the independent spin engine.

The comparison with the results from cyclic many-body thermal engines of the last section is intriguing. On the one hand, the $(n+2)/3$-fold enhancements coincides with the $(ns+1)/(s+1)$-fold enhancement of the {\it work extracted per cycle} of the cyclic collective engine with spins of dimension $s=1/2$. 
However, at low hot bath temperature, the cyclic collective engine extracts per cycle $n$ times less work than the cyclic independent engine whereas at low effective temperature both continuous collective and independent engines produce the same output power. 
 On the other hand, comparing directly the output powers of cyclic and continuous engines, the maximal enhancements for collective cyclic engines is a $n(ns+1)/(s+1)$-fold enhancement, much larger than the $(n+2)/3$-fold enhancement of the continuous collective engines.

Note that this manifest non-equivalence between continuous and cyclic collective engine is not in contradiction with the result of equivalence proved in \cite{Uzdin_2015} since the model of cyclic engine considered in \cite{bathinducedcoh,colheatcapacity} does not correspond to vanishing actions, regime where the equivalence holds. 

\subsubsection{Bath-induced coherences in continuous degenerate engines}\label{secdegenerate} 
A simple theoretical model to investigate the influence of degenerate working medium in continuous engine can be a two-level system with $N-1$ degenerate excited levels and a single ground state \cite{Gelbwaser_2015,Niedenzu_2015,Gelbwaser_2015b}.
Having an external drive of the form $\hbar\omega(t)H_0$, where $H_0$ stands for the dimensionless Hamiltonian, ensures that no coherence is generated from the driving. 

The role of coherences can be studied considering the situation where all $N-1$ transition dipoles are parallel, reminiscent of the multi-spin situations (see Sections \ref{seccyclicmb}, \ref{secmb}), so that the interaction with the thermal bath can naturally generate coherences between the excited states. As a comparison, one considers incoherent configurations consisting in non-parallel transition dipoles, which generates independent transitions and no coherences. 
It appears that at large effective temperature $T_{\rm eff}$ (the same as in the previous section), the fully coherent design can exhibit a $N/2$-fold enhancement with respect to the incoherent engine, while such enhancement disappears at low effective temperature \cite{Niedenzu_2015}. As in the previous section, this behaviour relies on the properties of the steady state populations and can also be seen as ``steady-state superradiance''. 
This can be interpreted and understood as a consequence of bath-induced horizontal coherences which alter the apparent temperatures and subsequently the steady state populations, as in Sect.~\ref{seccyclicmb} and Sect.~\ref{secmb}.

\subsubsection{Bath-induced coherences in autonomous machines}\label{secauto}
Effects of bath-induced coherences have also been studied in different models of autonomous refrigerators powered by a third thermal baths, sometimes called absorption refrigerators. 
One of the simplest and most studied absorption refrigerator consists in three qubits interacting via a three-body interaction with each thermal bath inducing transitions in only one of the qubits \cite{Kosloff_2014,Mitchison_2019}. 
When the baths are common to the three qubits, inducing transitions in other available resonant transitions (delocalised between the spins), the baths generate horizontal coherences between degenerate collective levels \cite{Manzano_2019}, which alters the heat flows and consequently the performances of the refrigerator. It is shown explicitly \cite{Manzano_2019} by comparison with a dephased refrigerator that such bath-induced coherences increase the cooling power for some ranges of the bath temperatures. The physics and the conclusions regarding the role of bath-induced coherences are similar to the one of the previous sections Sect.~\ref{seccyclicmb} and Sect.~\ref{secmb}.

However, building on the observation that coherences are not always beneficial, one can come up with specific designs emphasising that.
For instance, consider a four-level refrigerator where the cold and the work bath (the third bath playing the role of the battery) interact in phase with the transition denoted by $|1\ket \leftrightarrow |a\ket$ but in opposite phase with the quasi-degenerate transition $|1\ket \leftrightarrow |b\ket$ \cite{Kilgour_2018}. As a consequence, the bright state of the cold bath is orthogonal to the bright state of the work bath, which simply switches off the refrigerator. However, an additional dephasing bath inducing a transition between these two states can switch on the refrigerator but simultaneously leads to a loss of bath-induced coherences \cite{Kilgour_2018}. 
Although not having a clear experimental implementation (in relation to the dephasing between the bath interactions), it illustrates well the double role of coherences.


An other design of autonomous machine using a four-level working medium with two degenerate levels was also suggested to explore the role of bath-induced coherences \cite{Scully_2011,Svidzinsky_2012,Holubec_2018,Holubec_2019,Rahav_2012,Goswami_2013,Dorfman_2018}. 
Due to the level degeneracy, the baths induce horizontal coherences between such two degenerate levels. The underlying mechanism is the same as the generation of delocalised excitations described in Sect.~\ref{seccyclicmb} (but with ``delocalisation'' between levels instead of spins). 
The general conclusions are that while the efficiency is not affected, both the output power and the efficiency at maximum power can be increased depending on the bath temperatures, on the operation regime (work extraction or refrigeration), and on which levels are degenerate. 

Finally, one can wonder how quantum are such enhancements and if the presence of coherences is enough to claim quantumness? It can be shown \cite{Holubec_2018}, for instance in situations where bath-induced coherences are maximal, that the dynamics can be described thanks to a change of basis by a classical stochastic process, meaning that no coherences are needed to describe the dynamics. This tends to suggest that the above enhancements are not exclusively quantum. We come back briefly on these ideas in the conclusion. 
Additionally, it appears that bath-induced coherences affect the heat flow fluctuations. However, some studies point at a reduction of the fluctuations \cite{Rahav_2012} while others studies conclude the opposite \cite{Holubec_2019}, which would tend to relativise the gain of output power brought by bath-induced coherences. Note that the reason of this discord might be the location of the degeneracy, ground state \cite{Rahav_2012} versus excited state \cite{Holubec_2019}. 

\subsection{Coherences induced by interacting many-body working medium in autonomous machines} 
We finally mention briefly the situation where coherences are generated through interaction between the subsystems composing the working medium of autonomous machines. 
Coming back to the three-qubit absorption refrigerator of the last section, one can consider an additional strong interaction between the two qubits in contact respectively with the hot and work bath \cite{Du_2018}. Such strong coupling generates coherences between these two qubits. 
However, the overall effect is mixed: while the coherences increase the cooling power, they reduce the efficiency. 
Note that the efficiency considered in \cite{Du_2018} is somehow a partial efficiency since heat flows not related to the three-body interaction are left out. The relation between the total efficiency and the coherences is much less obvious.

Alternatively, the standard three-qubit refrigerator (without the above strong coupling between the hot and work qubits) can be analysed in the transient regime \cite{Mitchison_2015}. Cooling enhancement is observed and attributed to coherences within the three qubit ensemble emerging from their interaction. A related setup consisting of three harmonic oscillators evolving unitarily via three-body interaction is shown to exhibit similar transient cooling enhancement \cite{Nimmrichter_2017}. While such enhancement could be associated to the emergence of coherences as in the qubit case, it can be shown that a classical model can reproduce the same enhancements, casting doubt on the supposed quantum advantage. These are debated questions  still under investigation.

\section{Non-local coherences between working medium and bath}\label{secnl}
An other natural situation where coherences emerge is when the working medium is strongly coupled to the baths. 
Considering an Otto engine, the analysis can be made by comparing the performances of the strong coupling engine with the ones of a ``dephased'' engine \cite{Newman_2019}. This dephased engine is defined by cancelling, after each isochore, all the coherences contained in the extended working medium, which is composed of the original working medium --a two-level system-- and the reaction coordinates \cite{Newman_2017} (extra modes introduced to map the strong coupling dissipation to a Markovian dissipation). Thus, part of the cancelled coherences are non-local, meaning coherences present globally in the ensemble working medium plus bath, but not present locally in the working medium.  
%
%
 %
 
The conclusion is that such coherences have deleterious effects on both efficiency and output power. One could see this phenomenon as a different kind quantum friction, not generated by non-adiabatic coupling but by strong bath coupling. 
Beyond that, the coherences remaining within the original working medium due to strong bath coupling might also be detrimental per se, but provide a natural situation to investigate the impact coherences maintained throughout the cycle (instead of using incomplete isochore). 

 Finally, in the last section we saw that bath-induced coherences were beneficial, at least for some range of bath temperatures, while in this section it appears as detrimental. This is because the bath-induced coherences, beyond being vertical coherences, stem from residual correlations between working medium and bath whereas the bath-induced coherences considered in Sect.~\ref{secbathinduced} stems from indistinguishability of some degree of freedom. 
 Could it be more general, meaning that induced vertical coherences are always detrimental whereas horizontal ones can be harnessed to obtain positive effects?

 \section{Summary and final remarks}\label{concl}
 As mentioned in the introduction, the effects of coherences in thermal machines is a very intricate problems. Indeed, we saw lots of different effects in a large variety of situations. Still, we can identify some general tendencies.
Regarding coherences present in baths, we saw that horizontal coherences can be beneficial for the machine's efficiency. The amount of the enhancement can be easily assessed thanks to the apparent temperatures. By contrast, vertical coherences present in squeezed baths have no influences for the performances of the machines. More general vertical coherences can lead to work-like contributions and might have a positive impact although their experimental realisation appears as very challenging. 
The same conclusions are valid for work bath or batteries (for dispersive two-body coupling with the working medium) since the working medium only ``sees'' their apparent temperatures. However, the common underlying question is how such horizontal coherences can be created in the first place?
Positive effects from coherences accumulated in the battery after each cycle \cite{Watanabe_2017} bring interesting perspectives and deserve further attention.

Coherences induced in the working medium by non-adiabatic driving are always detrimental for power and efficiency 
unless such coherences are conserved thanks to incomplete isochores in order to interfere in the next driven stroke. Such effects were shown to be able to bring some positive effects but are still not well understood and should be more investigated and extended to include coherences induced by strong bath coupling. 
 Additionally, the impact of drive-induced coherences in continuous machines is not very well-known so that further studies would be welcome.
 Alternative strategies include extracting the drive-induced coherences to use it in other machines, or simply to avoid their generation 
 thanks to quantum lubrification or shortcut to adiabadicity techniques.

Bath-induced coherences can increase the output power of engines and refrigerators without affecting their efficiency. For some designs, this might be true only for certain ranges of the bath temperatures which depend on the details of the design, namely the energy level configuration of the working medium, its initial state, and wether the machine is a refrigerator or an engine. A general criteria encompassing all these aspects would be welcome for autonomous machines. 


  This brings the important ongoing debate about whether the presence of quantum coherences is enough to claim quantum advantages. There are several criteria. The most natural and mildest one is simply to compare the performances of the engine with and without coherences as for instance in \cite{Uzdin_2015,Klatzow_2019}. 
   A much more stringent criteria that we briefly mentioned in Sect.~\ref{secauto} is that quantum advantages can be claimed only if the dynamics of the quantum devices cannot be reproduced by any incoherent simulator \cite{Holubec_2018,Niedenzu_2018b,Gonzalez_2019}. This criteria goes beyond the relatively accepted notion of ``fair comparison'', stating that quantum and classical devices should be compared based on the same resources. Such conflict of criteria appeared for instance when the performances of an engine using an ensemble of $n$ spins $s$ is compared with an engine using a single spin $ns$. This review is not defending a specific criterion  since further investigations are needed to reach more compelling arguments, but  instead mentions them to promote this important debate. 
We hope this brief review will help to clarify the role of coherences in thermal machines and also point at promising directions toward practical and tractable realisations of quantum coherence-enhanced thermal machines. 

\section*{Acknowledgments}
This work is based upon research supported by the South African Research Chair Initiative, Grant No. UID 64812 of the Department of Science and Technology of the Republic of South Africa and National Research Foundation of the Republic of South Africa.


\begin{thebibliography}{}
\bibitem{Carnot} S. Carnot, \textit{R{\'e}flections sur la Puissance Motrice du Feu et sur les Machines propres {\`a} D{\'e}velopper cette
Puissance}. Paris: Bachelier, (1824).
\bibitem{Scovil_1959} H. E. D. Scovil, E. O. Schulz-DuBois, Phys. Rev. Lett. {\bf 2}, 262 (1959).
\bibitem{Geusic_1967} J. Geusic, E. O. Schulz-Du Bois, H. E. D. Scovil, Phys. Rev. {\bf156}, 343 (1967).
\bibitem{Levy_2019} A. Levy and D. Gelbwaser-Klimovsky, \textit{Quantum Features and Signatures of Quantum Thermal Machines}. In: Binder F., Correa L., Gogolin C., Anders J., Adesso G. (eds) \textit{Thermodynamics in the Quantum Regime. Fundamental Theories of Physics}, vol 195. Springer, Cham (2018).
\bibitem{Quan_2007} H. T. Quan, Yu-xi Liu, C. P. Sun, and Franco Nori, Phys. Rev. E {\bf76}, 031105 (2007).
\bibitem{Kosloff_2014} R. Kosloff and A. Levy, Annu. Rev. Phys. Chem. {\bf65}:365-93 (2014).
\bibitem{Mitchison_2019} Mark T. Mitchison, Contemporary Physics {\bf60}, 164 (2019).
\bibitem{Gelbwaser_2014} D. Gelbwaser-Klimovsky and G. Kurizki, Phys. Rev. E {\bf90}, 022102 (2014).



\bibitem{Huang_2012} X. L. Huang, Tao Wang, and X. X. Yi, Phys. Rev. E {\bf 86}, 051105 (2012).
\bibitem{Abah_2014} O. Abah, E. Lutz, Euro. Phys. Lett. {\bf106}, 20001 (2014).
\bibitem{RoBnagel_2014} J. Ro{\ss}nagel, O. Abah, F. Schmidt-Kaler, K. Singer, and E. Lutz, Phys. Rev. Lett. {\bf112}, 030602 (2014).
\bibitem{Manzano_2016} G. Manzano, F. Galve, R. Zambrini, and J. M. R. Parrondo, Phys. Rev. E {\bf93}, 052120 (2016). 
\bibitem{Niedenzu_2016} W. Niedenzu, D. Gelbwaser-Klimovsky, A. G. Kofman and G. Kurizki, New J. Phys. {\bf18}, 083012 (2016). 
\bibitem{Niedenzu_2018} W. Niedenzu, V. Mukherjee, A. Ghosh, A. G. Kofman, and G. Kurizki, Nat. Comm. {\bf 9}:165 (2018).
\bibitem{Xiao_2018} B. Xiao, R. Li, Phys. Lett. A {\bf382}, 3051-3057 (2018).
\bibitem{Alicki_2015} R. Alicki and D. Gelbwaser-Klimovsky, New J. Phys. {\bf17}, 115012 (2015).
\bibitem{Agarwalla_2017} B. Kumar Agarwalla, J.-H. Jiang, and D. Segal, Phys. Rev. B {\bf  96}, 104304 (2017).
\bibitem{Ferraro_2005} A. Ferraro, S. Olivares, M. Paris, \textit{Gaussian States in Quantum Information}, (Bibliopolis, Napoli, 2005).
\bibitem{CAbound} F. Curzon, B. Ahlborn, Am. J. Phys. {\bf43}:22-24 (1975).
\bibitem{Klaers_2017} J. Klaers, S. Faelt, A. Imamoglu, and E. Togan, Phys. Rev. X {\bf7}, 031044 (2017).
\bibitem{paperautonomous} C. L. Latune, I. Sinayskiy, and F. Petruccione, Sci. Rep. {\bf9}:3191 (2019).
\bibitem{Alicki_2014} R. Alicki, arXiv:1401.7865.
\bibitem{paperapptemp} C. L. Latune, I. Sinayskiy, F. Petruccione, Quantum Sci. Technol. {\bf 4} (2019) 025005.
\bibitem{heatflowreversal} C. L. Latune, I. Sinayskiy, F. Petruccione, Phys. Rev. Research {\bf1}, 033097 (2019).


\bibitem{Petruccione_Book} H. Breuer and F. Petruccione, \textit{Theory of Open Quantum Systems}, (Oxford, Oxford, 2002).

\bibitem{Correa_2014} L. A. Correa, J. P. Palao, D. Alonso, and G. Adesso, Sci. Rep. {\bf4}:3949 (2014).
\bibitem{Dag_2016} C. B. Da{\u g}, W. Niedenzu, {\"O}. E. M\"{u}stecapl\i o\v{g}lu, and G. Kurizki, Entropy {\bf18}, 244 (2016).
\bibitem{Rodrigues_2019} F. L. S. Rodrigues, G. De Chiara, M. Paternostro, G.T. Landi, Phys. Rev. Lett. {\bf123}, 140601 (2019).
\bibitem{Scully_2003} M. O. Scully, M. S. Zubairy, G. S. Agarwal, and H. Walther, Science {\bf299}, 862 (2003).
\bibitem{Turkpence_2016} D. T{\"u}rkpen{\c c}e and {\"O}. E. M\"{u}stecapl\i o\v{g}lu, Phys. Rev. E {\bf 93}, 012145 (2016).
\bibitem{Guff_2019} T. Guff, S. Daryanoosh, B. Q. Baragiola, A. Gilchrist, Phys. Rev. E {\bf100}, 032129 (2019).
\bibitem{Dillenschneider_2009} R. Dillenschneider and E. Lutz, EPL {\bf8}, 50003 (2009).
\bibitem{Liberato_2011} S. De Liberato and M. Ueda, Phys. Rev. E {\bf 84}, 051122 (2011).
\bibitem{Li_2014} H. Li, J. Zou, W.-L. Yu, {\it et al.}, 
Phys. Rev. E {\bf 89}, 052132 (2014).
\bibitem{Filipowicz_1986} P. Filipowicz, J. Javanainen and P. Meystre, Phys. Rev. A {\bf34}, 3077 (1986).
\bibitem{Milburn_1987} G.J. Milburn, Phys. Rev. A {\bf36}, 744 (1987).
\bibitem{Alicki_1987} R. Alicki and K. Lendi, \textit{Quantum Dynamical Semigroups and Applications} (Lecture Notes in Physics vol 717) (Berlin: Springer, 1987).


\bibitem{Gelbwaser_2015} D. G.-Klimovsky, W. Niedenzu, P. Brumer, and G. Kurizki, Sci. Rep. {\bf 5}:14413 (2015).

\bibitem{Doyeux_2016} P. Doyeux, B. Leggio, R. Messina, and M. Antezza, Phys. Rev. E {\bf93}, 022134 (2016).
\bibitem{Watanabe_2017} G. Watanabe, B. P. Venkatesh, P. Talkner, and A. del Campo, Phys. Rev. Lett. {\bf118}, 050601 (2017).
\bibitem{Watanabe_2020} G. Watanabe, B.P. Venkatesh, P. Talkner, M.-J. Hwang, and A. del Campo,  Phys. Rev. Lett. {\bf 124}, 210603 (2020).


\bibitem{Teufel_2003} S. Teufel, \textit{Adiabatic Perturbation Theory in Quantum Dynamics}, (Berlin:Springer, 2003).
\bibitem{Allahverdyan_2005} A. E. Allahverdyan and Th. M. Nieuwenhuizen, Phys. Rev. E {\bf71}, 046107 (2005).
\bibitem{Albash_2012} T. Albash, S. Boixo, D. A Lidar and P. Zanardi, New J. Phys. {\bf14}, (2012) 123016.
\bibitem{Plastina_2014} F. Plastina, A. Alecce, T. J. G. Apollaro, {\it et al.},
Phys. Rev. Lett. {\bf 113}, 260601 (2014). 
\bibitem{Korzekwa_2016} K. Korzekwa, M. Lostaglio, J. Oppenheim, D. Jennings, New J. Phys. {\bf18}, 023045 (2016).

\bibitem{negentropyprod} C. L. Latune, I. Sinayskiy, F. Petruccione, Phys. Rev. A {\bf102}, 042220 (2020).  



\bibitem{Kosloff_2002} R. Kosloff and T. Feldmann, Phys. Rev. E {\bf65}, 055102(R)
(2002).
\bibitem{Feldmann_2003} T. Feldmann and R. Kosloff, Phys. Rev. E {\bf68}, 016101 (2003).
\bibitem{Feldmann_2004} T. Feldmann and R. Kosloff, Phys. Rev. E {\bf70}, 046110 (2004).

\bibitem{Brandner_2016} K. Brandner and U. Seifert, Phys. Rev. E {\bf 93}, 062134 (2016).
\bibitem{Brandner_2017} K. Brandner, M. Bauer, and U. Seifert, Phys. Rev. Lett. {\bf 119}, 170602 (2017).
\bibitem{Brandner_2020} K. Brandner and K. Saito, Phys. Rev. Lett. {\bf 124}, 040602 (2020).
\bibitem{Feldmann_2006} T. Feldmann and R. Kosloff, Phys. Rev. E {\bf73}, 025107(R) (2006).




\bibitem{Berry_2009} M. V. Berry, J. Phys. A {\bf42}, 365303 (2009).
\bibitem{Chen_2010} X. Chen, A. Ruschhaupt, S. Schmidt, {\it et al.},
Phys. Rev. Lett. {\bf104}, 063002 (2010).
\bibitem{Deng_2018} S. Deng, A. Chenu, P. Diao, {\it et al.}, Science Advances 4, eaar5909 (2018).
\bibitem{delCampo_2014} A. del Campo, J. Goold and M. Paternostro 2014 Sci. Rep. {\bf4}: 6208 (2014).



\bibitem{Deng_2013} J. Deng, Q.-h. Wang, Z. Liu, P. H{\"a}nggi and J. Gong, Phys. Rev. E {\bf88}, 062122 (2013).

\bibitem{Beau_2016} M. Beau, J. Jaramillo and A. del Campo, Entropy {\bf18}, 168 (2016). 


\bibitem{Abah_2019} O. Abah and M. Paternostro, Phys. Rev. E {\bf99}, 022110 (2019).

\bibitem{Hartmann_2020} A. Hartmann, V. Mukherjee, W. Niedenzu, and W. Lechner, Phys. Rev. R. {\bf 2}, 023145 (2020).


\bibitem{Dann_2020} R. Dann and R. Kosloff, New J. Phys. {\bf 22}, 013055 (2020).
\bibitem{Campbell_2017} S. Campbell, S. Deffner, Phys. Rev. Lett. {\bf118}, 100601 (2017).
\bibitem{Abah_2019b} O. Abah, R. Puebla, A. Kiely, G. De Chiara, M. Paternostro and S. Campbell, New J. Phys. {\bf21}, 103048 (2019).
\bibitem{Abah_2018} O. Abah and E. Lutz, Phys. Rev. E {\bf98}, 032121 (2018).
\bibitem{Abah_2019c} O. Abah, M. Paternostro, and E. Lutz, Phys. Rev. Research {\bf2}, 023120 (2020).

\bibitem{Uzdin_2016} R. Uzdin, Phys. Rev. Applied {\bf6}, 024004 (2016).


\bibitem{Feldmann_2012} T. Feldmann and R. Kosloff, Phys. Rev. E {\bf85}, 051114 (2012).
\bibitem{Camati_2019} P. A. Camati, J. F. G. Santos, and R. M. Serra, Phys. Rev. A {\bf99}, 062103 (2019).
\bibitem{Altintas_2015} F. Altintas, A {\"U} . C. Hardal, and {\"O}. E. M\"{u}stecapl\i o\v{g}lu, Phys. Rev. A {\bf 91}, 023816 (2015).




\bibitem{Uzdin_2015} R. Uzdin, A. Levy, and R. Kosloff, Phys. Rev. X {\bf 5}, 031044 (2015).
\bibitem{Klatzow_2019} J. Klatzow, J. N. Becker, P. M. Ledingham, {\it et al.},
Phys. Rev. Lett. {\bf122}, 110601 (2019).

\bibitem{Scully_2011} M. O. Scully, K. R. Chapin, K. E. Dorfman, M. B. Kim, and A. Svidzinsky, Proc. Natl.
Acad. Sci. USA {\bf108} (37), (2011) 15097.
\bibitem{Svidzinsky_2012} A. A. Svidzinsky, K. E. Dorfman, M. O. Scully, Coherent Optical Phenomena {\bf1}, 7 (2012).



\bibitem{Kloc_2019} M. Kloc, P. Cejnar, and G. Schaller, Phys. Rev. E {\bf100}, (2019) 042126.
\bibitem{colheatcapacity} C. L. Latune, I. Sinayskiy, F. Petruccione, New J. Phys. {\bf 22}, 083049 (2020).
\bibitem{Gross_1982} M. Gross and S. Haroche, Physics Reports  {\bf 93}, 301-396 (1982).
\bibitem{bathinducedcoh}C. L. Latune, I. Sinayskiy, F. Petruccione, Phys. Rev. Research {\bf1}, 033192 (2019).

\bibitem{bathinducedcohTLS} C. L. Latune, I. Sinayskiy, F. Petruccione, Phys. Rev. A {\bf99}, 052105 (2019).
\bibitem{Kim_2002} M. S. Kim, J. Lee, D. Ahn, and P. L. Knight, Phys. Rev. A {\bf 65}, 040101 (2002).
\bibitem{Plenio_2002} M. B. Plenio and S. F. Huelga, Phys. Rev. Lett. {\bf 88}, 197901 (2002).
\bibitem{Benatti_2003} F. Benatti, R. Floreanini, and M. Piani, Phys. Rev. Lett. {\bf 91}, 070402 (2003). 
\bibitem{Niedenzu_2018b} W. Niedenzu and G. Kurizki, New J. Phys. {\bf20}, (2018) 113038.
\bibitem{Holubec_2018} V. Holubec , and T. Novotn{\'y}, J. Low Temp. Phys. {\bf192}, 147 (2018).
\bibitem{Holubec_2019} V. Holubec , and T. Novotn{\'y}, J. Chem. Phys. {\bf151}, 044108 (2019).
\bibitem{Niedenzu_2015} W. Niedenzu, D. Gelbwaser-Klimovsky, and G. Kurizki, Phys. Rev. E {\bf 92}, (2015) 042123. 
\bibitem{Gelbwaser_2015b} D. Gelbwaser-Klimovsky, W. Niedenzu, and G. Kurizki, Adv. At., Mol., Opt. Phys. {\bf64}, 329 (2015).






\bibitem{Manzano_2019} G. Manzano, G.-L. Giorgi, R. Fazio, and R. Zambrini, New J. Phys. {\bf21}, 123026 (2019).




\bibitem{Kilgour_2018} M. Kilgour and D. Segal, Phys. Rev. E {\bf98}, 012117 (2018).
\bibitem{Rahav_2012} S. Rahav, U. Harbola, and S. Mukamel, Phys. Rev. A {\bf 86}, 043843 (2012).
\bibitem{Goswami_2013} H. P. Goswami and U. Harbola, Phys. Rev. A {\bf88}, 013842 (2013).
\bibitem{Dorfman_2018} K. E. Dorfman, D. Xu, and J. Cao, Phys. Rev. E {\bf97}, 042120 (2018).

\bibitem{Du_2018} J.-Y. Du and F.-L. Zhang, New J. Phys. {\bf20}, 063005 (2018).
\bibitem{Mitchison_2015} M. T. Mitchison, M. P. Woods, J. Prior and M. Huber, New J. Phys. {\bf17}, 115013 (2015).
\bibitem{Nimmrichter_2017} S. Nimmrichter, J. Dai, A. Roulet, and V. Scarani, Quantum {\bf1}, 37 (2017).







\bibitem{Newman_2019} D. Newman, F. Mintert, and A. Nazir, Phys. Rev. E {\bf101}, 052129 (2020).
\bibitem{Newman_2017} D. Newman, F. Mintert, and A. Nazir, Phys. Rev. E {\bf95}, 032139 (2017). 


\bibitem{Gonzalez_2019} J. O. Gonz{\'a}lez, J. P. Palao, D. Alonso, and L. A. Correa, Phys. Rev E {\bf 99}, 062102 (2019). 




\end{thebibliography}
\end{document}